# Spontaneous Mirror Symmetry Breaking in the Limited Enantioselective Autocatalysis Model: Abyssal Hydrothermal Vents as Scenario for the Emergence of Chirality in Prebiotic Chemistry

Josep M. Ribó,[*a,b] Joaquim Crusats,[a,b] Zoubir El-Hachemi,[a,b] Albert Moyano,[a] Celia Blanco[c] and David Hochberg[c]

a) Department of Organic Chemistry, University of Barcelona. c. Martí i Franquès 1, 08028-Barcelona, Catalonia, Spain. Emails; jmribo@ub.edu: j.crusats@ub.edu: zelhachemi@ub.edu: amoyano@ub.edu

b) Institute of Cosmos Science, University of Barcelona (IEEC-UB), c. Martí i Franquès 1, 08028-Barcelona, Catalonia, Spain.

c) Centro de Astrobiología (CSIC-INTA), Ctra. Ajalvir Km. 4, 28850-Torrejón de Ardoz, Madrid, Spain. Emails; hochbergd@cab.inta-csic.es: blancodtc@cab.inta-csic.es

# ABSTRACT

The emergence of chirality in enantioselective autocatalysis for compounds unable to transform according to the Frank-like reaction network is discussed with respect to the controversial limited enantioselectivity (LES) model composed of coupled enantioselective and non-enantioselective autocatalyses. The LES model cannot lead to spontaneous mirror symmetry breaking (SMSB) either in closed systems with a homogeneous temperature distribution nor in closed systems with a stationary non-uniform temperature distribution. However, simulations of chemical kinetics in a two-compartment model demonstrate that SMSB may occur if both autocatalytic reactions are spatially separated at different temperatures in different compartments but coupled under the action of a continous internal flow. In such conditions the system can evolve, for certain reaction and system parameters, towards a chiral stationary state, i.e., the system is able to reach a bifurcation point leading to SMSB. Numerical simulations using reasonable chemical parameters suggest that an adequate scenario for such a SMSB would be that of abyssal hydrothermal vents, by virtue of the typical temperature gradients found there and the role of inorganic solids mediating chemical reactions in an enzyme-like role.

# INTRODUCTION

The origin of homochirality in biomolecules and its role in living systems are central points for understanding chemical evolution (Avalos et al., 2009; Guijarro and Yus, 2009; Luisi, 2006). Nowadays, there is an implicit agreement to attribute the advent of homochiral macromolecules (Avetisov V. and Goldanskii et al., 1996) to synthesis from homochiral mixtures or mixtures displaying a significant enantiomeric excess (*ee*) {*ee* (%) = ([L] − [D])/([L] + [D]) x 100} of small chiral building blocks (e.g. monomers) instead of from racemic mixtures and the subsequent spontaneous mirror symmetry breaking (SMSB) of the resulting racemic macromolecules. In the first scenario, in contrast to the second one, oligomers and macromolecules that are carriers of functional properties and/or enantioselective catalytic properties may appear at early stages of chemical evolution and are themselves able to evolve towards an increasing level of complexity. In such a scenario SMSB must occur in the early stages of abiotic chemical evolution. In this respect, there is a lack of results, supporting the possibility of SMSB during the primordial stages of abiotic chemistry, for example, concerning the topics of the abiotic synthesis of amino acids and of primitive metabolic cycles coming from the seminal reports of (Miller, 1953) and (Wächterhaüser, 1988 and 1990). This is probably due to the fact that classical organic synthesis considers absolute asymmetric synthesis as a Utopian task not worthy of an experimental research effort. However, recent synthetic reports have found SMSB in chemical systems either in organic reactions (Mauksch et al., 2007; Soai et al. 1995; Soai and Kawasaki, 2008) or in phase transitions, e.g. (Viedma, 2005; Noorduin et al. 2009; El-Hachemi et al., 2011). The common feature shared by all these reports is the presence of an enantioselective autocatalysis reaction.[1]

Autocatalytic reactions are rare, but the ongoing developments devoted to the research of the Soai reaction (Soai and Kawasaki, 2007) have taught us just how necessary the detailed experimental study is of a specific enantioselective reaction in order to relate it with a theoretical model that convert autocatalysis into a bifurcating SMSB reaction (i.e., stochastic chiral signs of the final products in the absence of any chiral polarization). Only after this is it possible to study the effect of extremely weak chiral polarizations, which act to transform the stochastic final chiral sign into a deterministic one, e.g. (Kawasaki et al., 2009). This last report is an experimental proof of the previously theoretically predicted effects of minute chiral polarizations on the bifurcation point of SMSB processes (Avetisov et al., 1987; Kondepudi, 1987; Kondepudi et al., 1985).

---

[1] With respect to the different complex reaction networks able to lead to a dynamic autocatalytic signature, see (Plasson et al. 2011).

As recently stated, the shroud of mystery surrounding the emergence of chirality in chemical evolution is vanishing (Weissbuch and Lahav, 2011). However, not too much is currently known about the necessary conditions and constraints for SMSB and about reasonable abiotic scenarios which would fulfill them. In our opinion, the recent advances on the topic suggest as a reasonable hypothesis that in chemical evolution chirality would emerge in different SMSB scenarios, which cooperatively would lead to chiral complex functional systems, resilient against racemization and able to exchange chiral compounds of homochiral sign. In such a framework, it is of interest to search for the general conditions and constraints on chemical transformations that can lead to SMSB and their relationship with reasonable abiotic scenarios.

Enantioselective autocatalysis (Plasson et al., 2007), when coupled to other reactions implying chiral catalysts, may lead to SMSB. In the Frank model (Frank, 1955) (Scheme 1), enantioselective autocatalysis amplifies the chiral fluctuations by the coupling with a "mutual inhibition" reaction ([a] in Scheme 1) that determines, by removing the chiral products/catalysts in racemic composition, the increase of the *ee* of the catalyst available for autocatalysis [2]. The Frank model leads to SMSB as a stationary state in a open system (exchange of matter and energy with the surroundings); however, it has been shown recently that in a closed system (only exchange of energy with the surroundings) may also lead to the emergence of chirality (Crusats et al., 2009; Rivera Islas et al., 2005). Notice that for a closed system the SMSB in the Frank model is a kinetically controlled emergence of chirality (i.e. the non-racemic state is doomed to be destroyed in the long term), however, the system is resistant to the eventual racemization that occurs much later than the time necessary to exhaust the initial achiral products: for relatively high exergonic transformations racemization occurs several orders of magnitude later and for highly exergonic transformations racemization occurs at geological time scales. Therefore, such a process following the Frank model can be of practical use in synthesis and in a prebiotic scenario could have acted as a source of enantiopure compounds for subsequent chemical events. A Frank model scenario explains the feasibility (Crusats et al., 2009) of the highly exergonic Soai reaction (Soai and Kawasaki, 2008) and probably also other results from less exergonic reactions (Mauksch et al., 2007). Notice that the Frank model of Scheme 1 is a reductionist approach to tease out the essential features, which can be fulfilled by other more complex reaction networks, for example by chiral polymerizations (Sandars, 2003; Hochberg, 2010), beyond those depicted in Scheme 1.

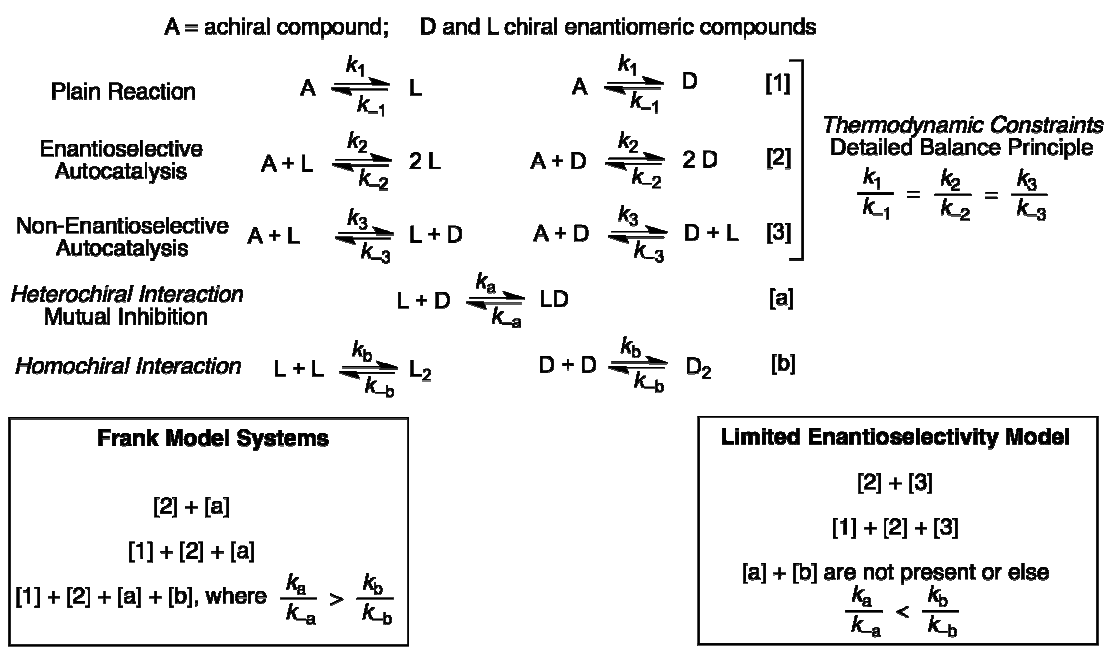

**Scheme 1.** Reaction networks of coupled enantioselective autocatalysis: the Frank model and the limited enantioselectivity (LES) model.

In a Frank-like network a necessary condition for SMSB is that the heterochiral interaction between products/catalysts is favored compared to the homochiral interaction (reactions [a] and [b] of Scheme 1, respectively). This is the case in the majority of chiral organic compounds, as can be inferred from the number of chiral compounds that crystallize as racemic crystals compared to those yielding a racemic mixture of enantiopure crystals (racemic conglomerates) (Jacques et al., 1981). However, this last group includes significant compounds in prebiotic chemistry, as for example, several amino acids. Therefore, it seems strange that the Frank model should be the only type of SMSB acting in chemical evolution. In this respect, the limited enantioselective (LES) model (Avetisov and Goldanskii, 1996), does not need nor depend on mutual interactions between the final products/catalysts (see Scheme 1). In the LES model the inverse reaction of the non-enantioselective autocatalysis (reaction [3] in Scheme 1) substitutes for the mutual inhibition reaction [a] in the Frank model (in Scheme 1). Previous reports had claimed SMSB in this model, but it has been demonstrated (Ribo and Hochberg, 2008) that it cannot occur in closed systems with a uniform temperature distribution (obviously also in the absence of selective energy input to only some species of the system, e.g. in the case of photo-stationary state): contradictory reports concerning this were consequence (Blackmond and Matar, 2008) of the use of a set of reaction rate constants which do not fulfill the principle of detailed balance.

A new scenario for SMSB in compounds for which the homochiral interactions are more favored than the heterochiral ones, i.e. for reactions that cannot follow a Frank-like model or scheme, is that of the dramatic experimental reports on the deracemization of racemic mixtures of crystals and on the crystallization from boiling solutions (Cintas and Viedma, 2011; El-Hachemi et al., 2011; Noorduin et al. 2008; Noorduin et al., 2009; Viedma, 2005; Wattis, 2011). This probably also occurs for other phase transitions, as indicated by a recent example on sublimations (Viedma et al., 2011). Such a SMSB is exhibited by achiral or fast racemizing compounds that crystallize as racemic conglomerates (the enantiopure crystal is more stable than the racemic crystal!). In spite of some controversy about the actual mechanisms acting in this SMSB, the experimental reports all agree that the final chiral state is a stationary state; for instance a mechano-stationary state in the case of wet grinding of racemic crystal mixtures (Noorduin et al. 2008; Noorduin et al., 2009; Viedma, 2005) and due to the presence of temperature gradients (system without uniform temperature distribution) in the case of deracemizations and crystallizations in boiling solutions (Cintas and Viedma, 2011; El-Hachemi et al., 2011). Only when the final conditions are those of constant energy input to the system and constant entropy production output to the surroundings, can the system be maintained away from the racemic state in a stationary manner. Note that this raises the question if the LES model could lead to SMSB as a stationary stable state when the system possesses a non-uniform temperature distribution, as is the case of the recent reports of the deracemization of racemic conglomerates.

Here we present results on the LES model with compartmentalized autocatalysis [2] and [3] in regions held at different temperatures (see Scheme 2). We discuss first the impossibility of SMSB in a system with a stable temperature gradient. Second, we show that SMSB in such a system is possible when the two autocatalyses [2] and [3] are compartmentalized in different temperature regions. Finally we speculate that a natural prebiotic scenario for such emergence of chirality is that of abyssal hydrothermal vents and volcanic plumes (Baross and Hoffman, 1985; Holm, 1992; Martin et al., 2008; Miller and Bada, 1988) which do have the adequate temperature gradients and contain solids, as for example clays, which have been proposed by several authors (Bernal, 1951; Brack, 2006; Cairns-Smith and Hartman, 1986; Ferris, 2005; Hazen, 2006)) as catalysts in the prebiotic synthesis of organic compounds.

## MATERIAL AND METHODS

Simulations were performed by numerical integration of the differential rate equations of all chemical species according to rate-equation theory as applied in chemical kinetics (mean field assumption). The concentration units are mol L$^{-1}$ and the different rate constants have the appropriate units to yield rate values in mol s$^{-1}$ unit. See in Results an example of the set of equations that describe the LES model in the two-compartment system. Numerical integration was performed with the Mathematica® program package. The results were monitored and verified to assure that the total system mass remained constant in time. For a set of parameters corresponding to the system at, or very near to, the bifurcation point the numerical integration is highly sensitive to minute differences between the reaction parameters, so that the inherent numerical noise of the calculations suffices to bifurcate the system towards a chiral outcome. In our simulations we have suppressed this computational noise, arising from round-off errors, by setting a high numerical precision of the input parameters (500 significant decimal digits and exact number representation of the reaction rates and the initial concentration values (for example "1 + 1·10$^{-2}$" instead of "1.01" or "1. + 1·10$^{-10}$" or "1 + 1.·10$^{-2}$"). Integration methods of "StiffnessSwitching" and "WorkingPrecision" of 60 were used in the calculations. The fluctuations of chirality able to change the racemic output to a chiral one were simulated by using an initial *ee* of catalysts lower than that expected from the statistical fluctuations around the ideal racemic composition (Mills, 1932; Mislow, 2003), i.e *ee* (%) < 67.43 x ($N^{-0.5}$), where $N$ is the number of chiral molecules. The numerical integration was run between 0 s to 1 x 10$^{17}$ s, i.e. from 0 s to 3.2 x 10$^{10}$ years. Using these integration methods, and in the metastable bifurcation region the strict absence of any initial enantiomeric excess and chiral polarization ($k_{\pm iD} = k_{\pm iL}$) the numerical integration yields a final racemic state. On the other hand, by using an initial *ee* slightly below the statistical fluctuations of the idealized racemic composition or extremely low differences between the enantiomeric transition states and the initial and final states [$k_{\pm iD} \neq k_{\pm iL}$; also for the exceedingly miniscule values such as those expected for the energy differences originating in the violation of electroweak parity (PVED)] then a chiral final state of high *ee* was obtained; furthermore, a correlation between the chiral signs of the initial "induction" and the final chiral sign of the *ee* was always obtained. In the case of reaction parameters leading to a racemic final state, this was confirmed by changing the initial conditions to high *ee* (same total concentration of chemical species): in the case of a racemic solution the final state was always racemic independently of the initial *ee*.

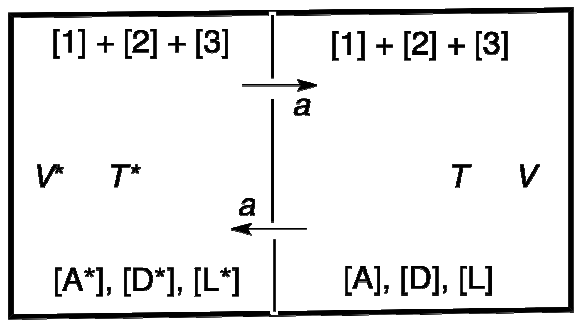

**Scheme 2.** LES model reaction network, transformations [1] + [2] + [3] (see text), in a two compartment system of volumes (*V* and *V\**) at different temperatures (*T* and *T\**) and with the exchange of solution (*a* L s$^{-1}$).

## RESULTS

The basic LES reaction network is formed by the following three transformations: [1] Non-catalyzed production of chiral compounds from an achiral substrate; [2] Enantioselective autocatalytic production and; [3] Non-enantioselective autocatalytic production.

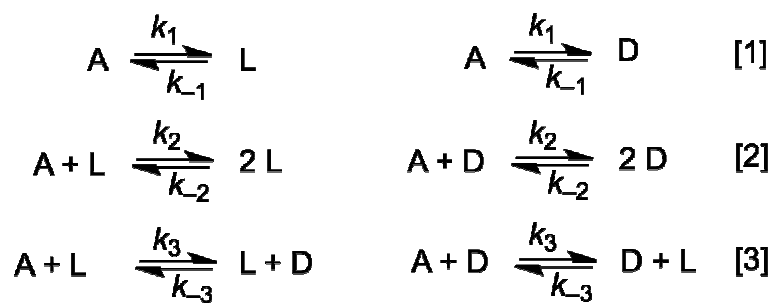

In these three transformations and for a given temperature *T* the following equation must be obeyed,

$$\frac{k_i}{k_{-i}} = K(T) \qquad (1)$$

where $K$ is the equilibrium constant of the transformation from A to D/L in [1]-[3]: the $K$ value depends on the temperature.

Because of the constraint (1) the LES reaction network can only lead to a racemic final equilibrium state (Ribo and Hochberg, 2009). The condition for SMSB in LES may occur for reaction parameters that do not fulfill equation (1). However, in equilibrium conditions the equation (1) must be fulfilled by virtue of the principle of detailed balance (microreversibility) (Prigogine and Kondepudi, 1998). SMSB in LES could occur when, in addition to a fast enantioselective autocatalysis [2] compared to the uncatalyzed reaction [1] ($k_2 >> k_1$ and $k_3 < k_2$), the condition $g < g_{crit}$ where $g = k_{-2}/k_{-3}$ and $g_{crit} \approx (1 - k_3/k_2)/(1 + 3\,k_3/k_2)$ holds [for details see (Ribo and Hochberg, 2009)]. Obviously this cannot be achieved when the thermodynamic constraint of equation (1) is fulfilled, because if $k_2 > k_3$ ($1/g = k_3/k_2$), then $k_{-2}$ is necessarily always greater than $k_{-3}$, i.e., $g$ is always greater than $g_{crit}$. This is the situation for a closed system with a uniform temperature distribution. However, we were intrigued by the possibility that SMSB might occur in a LES reaction network in a system with a stationary temperature gradient so arranged that the backward reaction of [3] in one region is faster than the backward reaction of [2] in the other region of the system.

*Limited Enantioselectivity (LES) in a two-compartment system at different temperatures.* In order to simulate a system with a stable temperature gradient we approximate by a virtual closed system of two compartments, held at different fixed temperatures, and coupled by a constant internal flow of solution ($a$ ; see Scheme 2) in both directions, assuming perfect uniform mixing conditions and fixed temperatures in both compartments (the concentration and reaction rate values in the compartment at the higher temperature are labeled by an asterisk; see Scheme 2). The corresponding differential equations of the rate equation model ([1] + [2] + [3]) of chemical kinetics in both compartments are:

$$\frac{d[\text{A}]}{dt} = -2k_1[\text{A}] - (k_2[\text{A}] + k_3[\text{A}] - k_{-1})([\text{D}] + [\text{L}]) + k_{-2}([\text{D}]^2 + [\text{L}]^2) + 2k_{-3}[\text{D}][\text{L}] + \frac{a}{V}([\text{A}^*] - [\text{A}]) \quad (2)$$

$$\frac{d[\text{L}]}{dt} = +k_1[\text{A}] + (k_2[\text{A}] - k_{-1})[\text{L}] - k_{-2}[\text{L}]^2 + k_3[\text{A}][\text{D}] - k_{-3}[\text{L}][\text{D}] + \frac{a}{V}([\text{L}^*] - [\text{L}]) \quad (3)$$

$$\frac{d[\text{D}]}{dt} = +k_1[\text{A}] + (k_2[\text{A}] - k_{-1})[\text{D}] - k_{-2}[\text{D}]^2 + k_3[\text{A}][\text{L}] - k_{-3}[\text{L}][\text{D}] + \frac{a}{V}([\text{D}^*] - [\text{D}]) \quad (4)$$

$$\frac{d[\text{A}^*]}{dt} = -k_1^*[\text{A}^*] - (k_2^*[\text{A}^*] + k_3^*[\text{A}^*] - k_{-1}^*)([\text{D}^*] + [\text{L}^*]) + k_{-2}^*([\text{D}^*]^2 + [\text{L}^*]^2) + 2k_{-3}^*([\text{D}^*][\text{L}^*]) + \frac{a}{V^*}([\text{A}] - [\text{A}^*]) \quad (5)$$

$$\frac{d[\text{L}^*]}{dt} = +k_1^*[\text{A}^*] + (k_2^*[\text{A}^*] - k_{-1}^*)[\text{L}^*] - k_{-2}^*[\text{L}^*]^2 + k_3^*[\text{A}^*][\text{D}^*] - k_{-3}^*[\text{L}^*][\text{D}^*] + \frac{a}{V^*}([\text{L}] - [\text{L}^*]) \quad (6)$$

$$\frac{d[\text{D}^*]}{dt} = +k_1^*[\text{A}^*] + (k_2^*[\text{A}^*] - k_{-1}^*)[\text{D}^*] - k_{-2}^*[\text{D}^*]^2 + k_3^*[\text{A}^*][\text{L}^*] - k_{-3}^*[\text{L}^*][\text{D}^*] + \frac{a}{V^*}([\text{D}] - [\text{D}^*]) \quad (7)$$

The system satisfies the conservation of total system mass (8).

$$V\left(\frac{d[\text{A}]}{dt} + \frac{d[\text{L}]}{dt} + \frac{d[\text{D}]}{dt}\right) + V^*\left(\frac{d[\text{A}^*]}{dt} + \frac{d[\text{L}^*]}{dt} + \frac{d[\text{D}^*]}{dt}\right) = 0 \quad (8)$$

Numerical integration simulations of (2)-(7) for many different parameter configurations obeying the constraint (1) lead, as expected, to a final racemic state. This is the consequence of the Onsager reciprocal relations (Prigogine and Kondepudi, 1998) which are the translation to non-equilibrium linear thermodynamics of the corresponding detailed balance principle at equilibrium conditions: notice that the rough approximations of the LES model of Scheme 2 (fixed and uniform temperature distribution and perfect mixing in each compartment) are analogous to those used to infer the Onsager reciprocal relations of reaction elementary steps and volumes at steady state conditions. In order to understand how the coupled reaction network of LES works in the two-compartment system, a discussion using simple chemical arguments is given here.

The reaction free energy $\Delta G°$ is the same in the three reactions [1], [2] and [3] and it is related to the free energy differences of the transition states ($\Delta G^{\ddagger}_{\pm i}$) by;

$$\Delta G^{\ddagger}_{-i} = \Delta G^{\ddagger}_{i} - \Delta G°. \qquad (9)$$

We can relate the reaction enthalpies and entropies with those of the transition state in the backward and forward reactions by a similar relationship;

$$\Delta H^{\ddagger}_{-i} = \Delta H^{\ddagger}_{i} - \Delta H°, \qquad (10)$$

$$\Delta S^{\ddagger}_{-i} = \Delta S^{\ddagger}_{i} - \Delta S°. \qquad (11)$$

The thermodynamic constraint (1), according to the Arrhenius equation and the Eyring theory of the transition state, taking into account (9)-(11), can be written as;

$$K(T) = k_i/k_{-i} = \exp[-\Delta G°/RT] = \exp[-\Delta H°/RT + \Delta S°/R] = \exp[-\Delta H°/RT]\exp[\text{constant}] \qquad (12)$$

this is in fact, the van't Hoff equation of the temperature dependence of $K$. The ratio between the K values at the temperatures T and T* ($k_i/k_{-i} = K_T$; $k^*_i/k^*_{-i} = K_{T*}$) as inferred from (12) is:

$$K_{T*}/K_T = \exp[-\Delta H°/R(1/T* - 1/T)] \qquad (14)$$

According to the Arrhenius equation for common chemical reactions ($\Delta H^{\ddagger} > 0$) the reaction rate increases when temperature increases and $K$, according to van't Hoff (12), in an exothermic reaction ($\Delta H° < 0$) it decreases when temperature increases and in a endothermic reaction it increases when temperature increases.

When we approximate the condition for SMSB in LES by a closed system [$k_{-3} > k_{-2}$ and $k_3 < k_2$ [(Ribo and Hochberg, 2009)]], as a reasonable conjecture for the two compartment system, it is obvious that it is not possible through a temperature change to achieve the condition of $k_{-3} < k_{-2}$ and $k_3 < k_2$ at T and $k^*_{-3} > k^*_{-2}$ and $k^*_3 > k^*_2$ at $T^*$ because of the dependence on the reaction enthalpy is the same for both the transformations [2] and [3].[2] Notice that the approximation of uniform mixing and uniform and constant temperatures within the compartments is equivalent to the differential elements applied in the demonstration of the Onsager reciprocal relations. Therefore in this model the flow exchange between both compartments will always lead to a racemic stationary state.

*Limited Enantioselectivity (LES) in a two compartment system at different temperatures and compartmentalized autocatalysis* [2] *and* [3]. The condition $k^*_{-3} > k_{-2}$ and $k^*_3 < k_2$ is possible thermodynamically in a thought (*Gedankenexperiment*) experiment that assumes, in addition to the presence of a temperature gradient, the separation of the autocatalysis reactions in the different temperature regions of the system; the non-enantioselective autocatalysis confined to the compartment at higher temperature ($T^*$) whereas the enantioselective autocatalysis is localized within the compartment at the lower temperature ($T$). Numerical integration simulations demonstrate that, for specific values of the reaction network parameters (concentrations and rate constants at $T$ and $T^*$) and of the system parameters ($V$, $V^*$ and $a$), the final stable state can be a chiral stationary state of high *ee*. This occurs when the autocatalysis [2] is more effective than the uncatalyzed plain reaction [1] ($k_2 \gg k_1$ and $k^*_2 \gg k^*_1$) and $k^*_{-3} > k_{-2}$ while $k^*_3 < k_2$. However, these conditions in the two-compartment system are probably necessary but not sufficient (see discussion below). Notice that the compartmentalization of [2] and [3] without any temperature difference between compartments, cannot lead to SMSB because of the constraint (1).

Figure 1 displays characteristic examples of such simulations obtained by numerical integration (Material and Methods) of the differential equations (2)-(7) (modified assuming the presence or absence of the transformations [2] or [3] in each compartment). In the next section we discuss the possible natural scenarios for this compartmentalized LES model, but the principal characteristics of such a system to yield SMSB are discussed here. Reasonable reaction parameters were used assuming a very slow uncatalyzed reaction [1] and good autocatalysis [2] and [3]. In the case of SMSB the evolution of the species concentrations can be qualitatively described to be composed by three stages; (1) The first stage is the evolution of the concentrations of A, D and L to equalize their concentrations in both compartments (not shown in the examples of Figs. 1-3, because of the equal initial concentrations in both compartments) that is faster than the; (2) the second stage is the conversion of the species towards the racemic steady state, but being this unstable a very small $ee \neq 0$ is mantained until the SMSB occurs; (3) finally, the amplification of this persistently small $ee$ to a high final $ee$ in the case of SMSB (bifurcation), or else racemization in the case of a racemic final state. When the internal flow value $a$ is very small, then the concentration equalization between both compartments occurs later than the transformation of the species, and then racemization dominates over the enantioselective amplification for any initial $ee$.

---

[2] We assume that anomalies leading to non-linear Arrhenius plots and to transition state positive enthalpies do not take place (Aquilanti et al., 2010, Olivenberg et al., 1995).

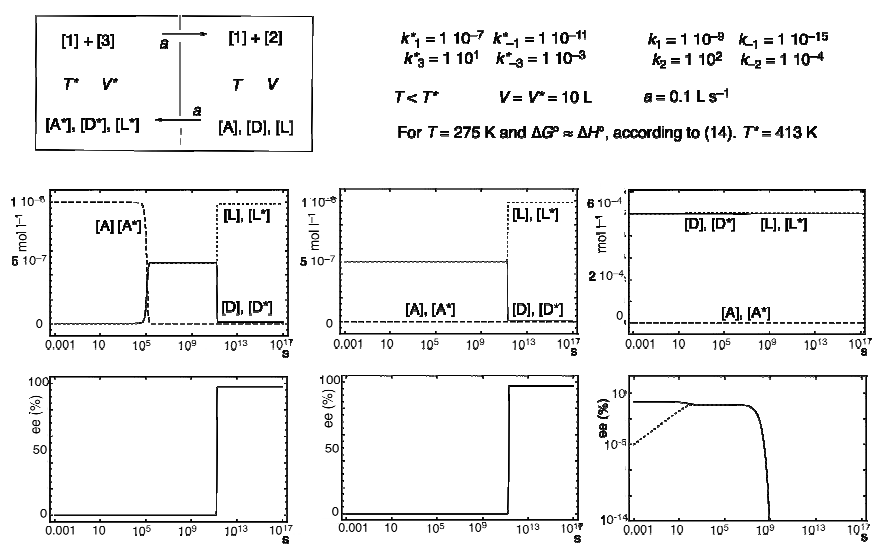

**Figure 1.** Simulations of the LES model in two compartments at different temperature ($T$ and $T^*$) and enantioselective autocatalysis ($T$) separated from non-enantioselective autocatalysis under the condition $k^*_{-3} > k_{-2}$ and $k^*_3 < k_2$. In order to reproduce chiral fluctuations from the ideal racemic composition, an initial $ee$ below the statistical deviation from the ideal racemic composition was assumed. *Left Column:* Low reactant concentration and initial conditions of transformation of A into L/D; $[A]_0 = 1 \times 10^{-6}$: $[L]_0 = 1 \times 10^{-11}$: $[D]_0 = 1 \times 10^{-11}$: $[A^*]_0 = 1 \times 10^{-6}$: $[L^*]_0 = 1 \times 10^{-11} + 1 \times 10^{-21}$: $[D^*]_0 = 1 \times 10^{-11}$. *Middle Column:* Low concentrations and starting conditions of equilibrium concentrations between A and D/L; $[A]_0 = 1 \times 10^{-11}$: $[L]_0 = 5 \times 10^{-7}$: $[D]_0 = 5 \times 10^{-7}$: $[A^*]_0 = 1 \times 10^{-11}$: $[L^*]_0 = 5 \times 10^{-7} + 1 \times 10^{-20}$: $[D^*]_0 = 5 \times 10^{-7}$. *Right Column:* Similar conditions as in the middle column but at higher total concentration lead to a final racemic state in spite of a significant initial $ee$ (continuous line compartment at $T^*$, doted line compartment at $T$); $[A]_0 = 1 \times 10^{-10}$: $[L]_0 = 5 \times 10^{-4}$: $[D]_0 = 5 \times 10^{-4}$: $[A^*]_0 = 1 \times 10^{-10}$: $[L^*]_0 = 5 \times 10^{-4} + 1 \times 10^{-5}$: $[D^*]_0 = 5 \times 10^{-4}$.

The final state, whether racemic or chiral, for a given total concentration is the same for any initial ratio between A, D and L, i.e., the final state is, although stationary, the most stable stationary state of the system: the racemic state is achieved also for high initial $ee$ and the chiral state can be achieved [in the presence of $ee$ far below the values of the statistical fluctuations (Mills, 1932; Mislow, 2003)] by the transformation from A to D and L (Fig.1 left column) as well from the "racemic" equilibrium concentrations (Fig.1 middle column).[3] This is in contrast with the kinetically controlled emergence of chirality in the Frank model in closed systems (Crusats et al., 2009) that depends on the initial ratio between chemical species. The racemic final state corresponds in fact to the mean of the equilibrium concentrations in both compartments. A chiral final state shows the same species concentration in both compartments but $[D]/[A] \neq [L]/[A]$ where $[A]$ is smaller than that of the corresponding racemic state, i.e., at the SMSB bifurcation not only does an increase of the concentration of one of the enantiomers take place, but also a decrease of $[A]$.

The LES system of Fig. 1 can lead to SMSB at a very low total concentration of the reactants ($[A]_o + [D]_o + [L]_o$). Furthermore, the achievement of the bifurcation leading to SMSB, when the condition $k^*_{-3} > k_{-2}$ and $k^*_3 < k_2$ is fulfilled, occurs below a critical value of the total concentration: above this critical concentration the system evolves to the racemic state and below it to the chiral state. Fig. 1 shows an example of how the increase of concentration values -from µM to mM order- leads from SMSB to a racemic final stationary state (compare middle column and right columns of Fig. 1). This critical concentration value depends also on the value of the exchanging flow *a* with respect to the compartment volumes *V* and *V*$^*$. For example, for the same conditions of the example of the right-hand column of Fig. 1 when *a* increases from 0.1 L s$^{-1}$ to 1.0 L s$^{-1}$, SMSB is achieved and the evolution with time of the chemical species is similar to those shown in the middle column. A phase and stability analysis of this system would in principle establish the mathematical dependence of all the system variables in an expression for its critical parameter: further work in this direction is in progress that will be published in a more specialized journal. Nevertheless, the numerical simulations performed strongly suggest that the effect of *a* on the critical concentration value is related to the rates of the forward reaction and backward reactions of the non-enantioselective autocatalysis. In this respect it is worth noting that the forward rate in the non-enantioselective autocatalysis [3] is a racemization, as well as the plain reaction, therefore the flow parameter *a* should be faster than such a racemization, furthermore, the backward reaction of the non-enantioselective autocatalysis tends to decrease the racemic composition that will be amplified by the enantioselective autocatalysis [2].

---

[3] In fact the definition here of the final racemic or chiral state is inferred from the numerical integrations when any low initial $ee$ leads to a final chiral state or any high initial $ee$ leads to an ideal racemic mixture.

In a SMSB bifurcation scenario (Avetisov et al., 1987; Kondepudi, 1987; Kondepudi et al. 1987) very low chiral polarizations are able to lead deterministically to one of the two final chiral states. Fig. 2 (left section) shows examples of this. SMSB is obtained by an enantioselective energy difference between the transition states, that would be the consequence of a chiral polarization. SMSB is also originated by the simulation of an intrinsic energy difference between the enantiomers, i.e. between the stabilities of L and D, on the order of the values expected from the effect of the weak force violation of parity (PVED) (right section Fig. 2).[4] It is also worth noting that the simultaneous action of chiral fields and concentration fluctuations about the racemic composition may amplify the effect of minute chiral polarizations (Cruz et al. 2008; Kondepudi, 1087).

---

[4] The free energy differences corresponding to the enantioselective rate constants of Fig. 2 (left section) and enantiomeric differences on the free energy between enantiomers (Fig. 2 right

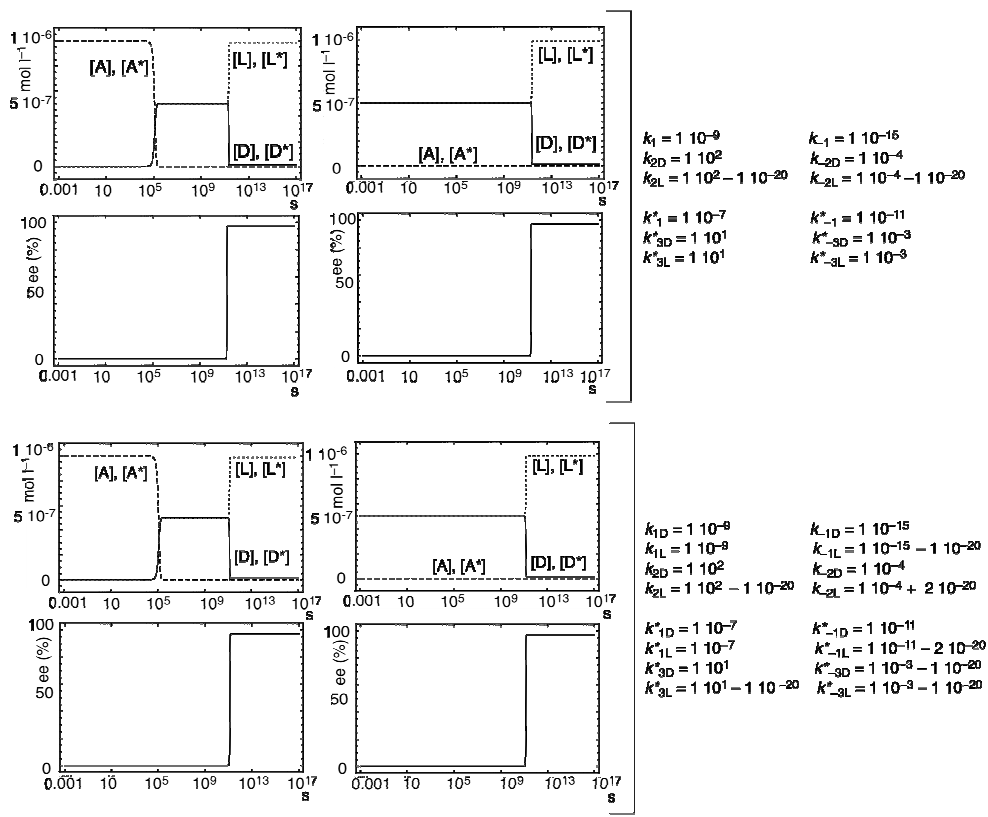

**Figure 2.** Simulations of the effect of low chiral polarization on the LES model of Fig. 1 ($V = V^* = 10$ l; $a = 0.1$ l s$^{-1}$). *First and second rows*; In [2], $\Delta G^{\ddagger}_D/RT - \Delta G^{\ddagger}_L/RT \approx 100$ fJ mol$^{-1}$. *Third and fourth rows;* In [1], [2] and [3] free energy difference between L and D on the order to simulate the expected energies differences by the parity violation (PVED ≈ 100 fJ mol$^{-1}$, with L more stable than D). *Initial conditions;* First row at third row; [A]$_o$ = 1 × 10$^{-6}$: [L]$_o$ = [D]$_o$ = 1 × 10$^{-11}$: [A$^*$]$_o$ = 1 × 10$^{-6}$: [L$^*$]$_o$ = [D$^*$]$_o$ = 1 × 10$^{-6}$. Second and fourth row; [A]$_o$ = 1 × 10$^{-11}$: [L]$_o$ = [D]$_o$ = 5 × 10$^{-7}$: [A$^*$]$_o$ = 1 × 10$^{-11}$: [L$^*$]$_o$ = [D$^*$]$_o$ = 5 × 10$^{-7}$.

In summary, the simulations of this thought experiment show that for adequate reaction parameters (reaction rate constants and concentrations) and system parameters (material exchange flow, compartment volumes and temperatures) a bifurcation SMSB

---

section) correspond to energy values of ≈ 100 fJ cal mol$^{-1}$, i.e., are of the same order as the PVED values estimated for common chiral organic compounds (Bonner, 1999; Gottselig and Quack, 2005; MacDermott, A.J. 1995).

scenario can be achieved capable of generating chirality at very low concentrations of the reactants. Notice that the constrained LES model presented here is able to deracemize a very dilute "racemic" soup, i.e., a solution of organic compounds that would be previously obtained from other sources, for example brought to Earth by extraterrestrial bodies. A possible real scenario for such a LES system is outlined in the following section.

## DISCUSSION

### Abiotic scenario for SMSB in the LES model.

Reactions mediated/catalyzed by a solid matrix or surface are accepted scenarios in prebiotic chemistry (Brack, 2006; Cairns-Smith and Hartman, 1986; Guijarro and Yus, 2009; Wächterhäuser, 1988 and 1990). Therefore, a feasible approximation to this LES model to an actual scenario is that of aqueous solutions containing specific solids, which would act as mediators for the enantio- and the non-enantioselective autocatalyses, situated at separate regions and at different temperatures. Such a model (Scheme 3) assumes a solid/surface inductor ($P_a$) at $T$ for the enantioselective autocatalysis and a solid inductor ($P_b$) at $T^*$ for the non-enantioselective autocatalysis ($T^* > T$), where the transformation from A to D/L is exothermic. Notice that now the constraint (1) is transformed into the conditions (15) and (16),

$$\frac{k_1}{k_{-1}} = \frac{k_a k_2}{k_{-a} k_{-2}} \qquad (15)$$

$$\frac{k_1^*}{k_{-1}^*} = \frac{k_b^* k_3^*}{k_{-b}^* k_{-3}^*} \qquad (16)$$

which are related by the temperature dependence through equation (14).

Fig. 3 shows an example of SMSB in the synthesis of D/L from achiral A and in the deracemization of a final mixture of A and racemic D/L. Reaction and system parameters that lead to the bifurcation (SMSB) define by themselves the possible astrobiological scenario. Such a scenario is that of deep ocean hydrothermal vents. The examples of Figs. 1-3 demonstrate that SMSB occurs for a temperature gradient that fits with the characteristics of the hydrothermal vents of the "Lost City" type (Martin et al., 2008), which are considered to be examples of the hydrothermal activity at the ocean bottom during the Hadean/Eo-Archean eons. In this respect, the scenario discussed here should be placed in time at an early stage of the transformation between organic compounds prior to the formation of the primordial cells.

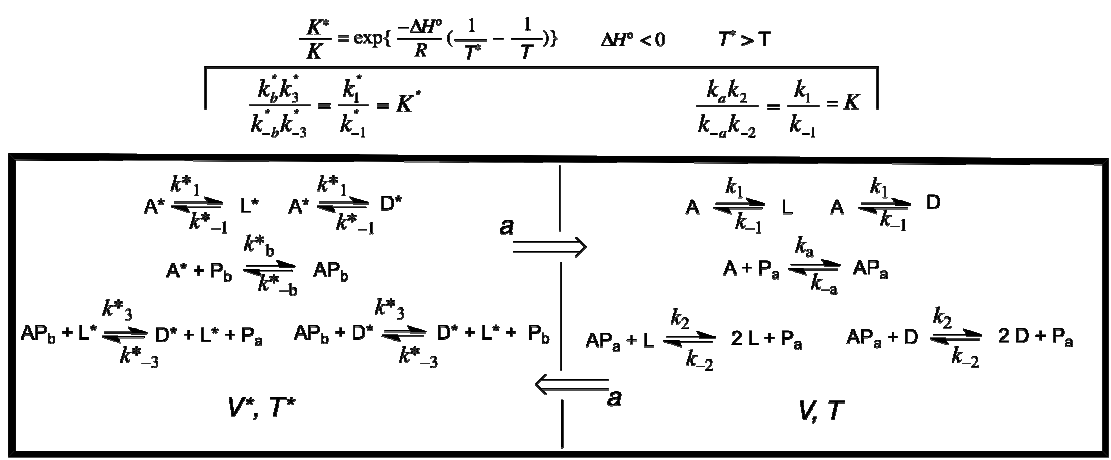

**Scheme 3.** Two compartment system at different temperatures and immobilized autocatalysis as a speculative model for the SMSB in abyssal hydrothermal vents (see text). Where A is achiral, D and L are enantiomeric compounds, $P_a$ and $P_b$ enzyme-like inmobilized mediators for the enantioselective- and non-enantioselective autocatalysis processes, respectively. The rest of system parameters are as defined in Fig. 1 (see text).

*Temperature gradients.* The conditions for bifurcation are achieved only for high temperature differences, which imply that the solvent can only remain in the liquid state under conditions of high pressure. The examples of Figs. 1-2 assuming $\Delta G° \approx \Delta H°$ ($\Delta H° \approx -7.6$ kcal mol$^{-1}$) yield SMSB at a temperature difference that can be found in abyssal hydrothermal vents of the Lost City type (for example, $T^* \approx 138°$ C; $T = 2°$ C) where water remains in the liquid or in supercritical state, that is, it can act as a solvent in the reactions [1]-[3].

180° C is a temperature value that should be taken as an upper limit for the thermal stability of organic compounds in solution. In this respect, Lost City type plumes are considered a better fit to the origin of life scenarios than the hotter black smoker vents (Martin et al., 2008).

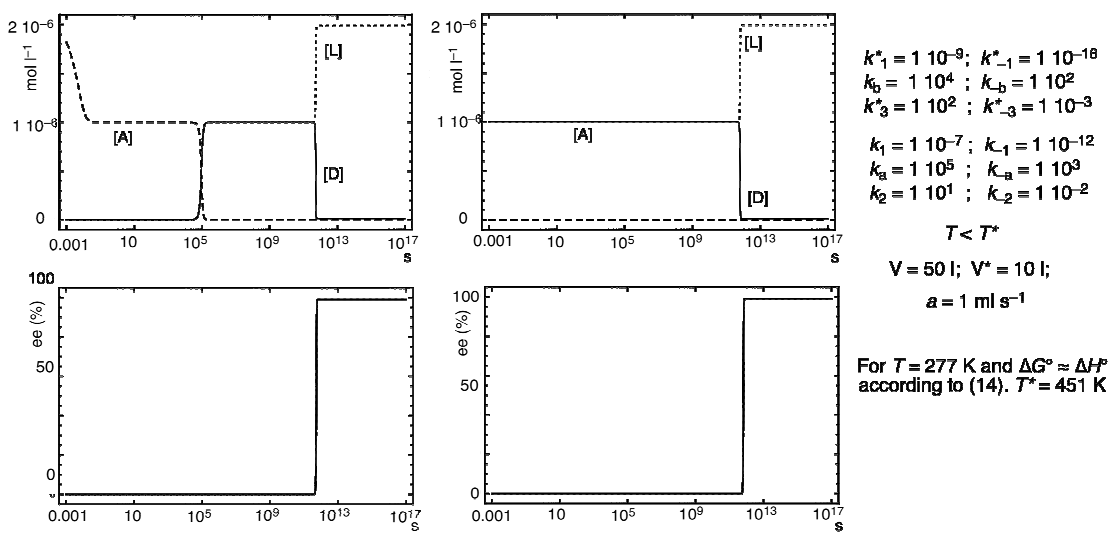

**Figure 3.** Simulations of SMSB in the LES model of Scheme 3 (evolution of concentrations in the compartment of higher volume at temperature T, but after the initial equalization of concentrations the evolution of the species concentration is the same in both compartments). *Left Column:* Low reactant concentration and initial conditions for the transformation of A to L and D; $[A]_o = 2 \times 10^{-6}$: $[L]_o = 1 \times 10^{-10} + 1 \times 10^{-20}$: $[D]_o = 1 \times 10^{-10}$: $[P_a]_o = 1 \times 10^{-2}$: $[AP_a]_o = 0$: $[A^*]_o = 2 \times 10^{-6}$: $[L^*]_o = 1 \times 10^{-10}$: $[D^*]_o = 1 \times 10^{-10}$: $[P_b]_o = 1 \times 10^{-2}$: $[AP_b]_o = 0$. *Right Column:* Low concentrations and starting conditions of equilibrium concentrations between A and D/L; $[A]_o = 1 \times 10^{-10}$: $[L]_o = 1 \times 10^{-6} + 1 \times 10^{-20}$: $[D]_o = 1 \times 10^{-6}$: $[P_a]_o = 1 \times 10^{-2}$: $[AP_a]_o = 0$: $[A^*]_o = 1 \times 10^{-10}$: $[L^*]_o = 1 \times 10^{-6}$: $[D^*]_o = 1 \times 10^{-6}$: $[P_b]_o = 1 \times 10^{-2}$: $[AP_b]_o = 0$.

*Solid and surface mediators for autocatalysis.* The possible role of phyllosilicates in abyssal hydrothermal vents as catalysts for the synthesis of organic compounds has been previously proposed (Martin et al., 2008); Williams et al. 2005). Furthermore, smectite-type clays in hydrothermal vents have been proposed not only to promote the development of abiotic organic compounds, but also to protect them against high temperatures (Williams et al., 2005).

The role of clays in abiotic chemistry was proposed many years ago and is an accepted scenario in the chemistry of the origin of life (Bernal, 1951; Brack, 2006; Cairns-Smith and Hartman, 1986). This is supported by the actual use of clay minerals as catalysts in organic synthesis (Adams and McCabe, 2006; Ferris, 2005; Son et al., 1998) and in stereoselective chemical processes: even low reproducible chiral recognition processes have been and are being sporadically reported (Guijarro and Yus, 2009). In this respect, some phyllosilicates, e.g. all natural kaolinites, are intrinsically chiral (Brigatti et al. 2006). It is worth noting that the role of clays in organic synthesis evoke a pseudo-enzymatic role in agreement with the role played by the solid/surface inductors of Scheme 3.

Here we assume a different stereospecific activity for the solid mediator at the two temperature sites. This specificity for enantio- or for non-enantioselective autocatalysis requieres a different chemical structure of the solid. This can be expected in an abyssal hydrothermal vent scenario that is even able to serpentinize olivine. The significant chemical and temperature gradients present in the Lost City vents gives a reasonable support for changes on the structure of the phyllosilicate at different sites of the vents: Lost City plumes show a pH gradient of 3-5 pH units between the inner walls (pH 9-10) and the surrounding bulk ocean. This and other chemical gradients (Martin et al, 2008), e.g. of cations, together with the temperature gradient could change the structure of the solid mediator to the extent of even changing the reaction mechanisms of the autocatalysis.

*The question of the time required for the SMSB.* The Figs. 1-3 show examples of SMSB which are difficult to be tested in applied chemistry. This not only because the experiments should be performed under high pressures in order to maintain fluid the media, but because of the long times required to achieve the final chiral stationary state: depending on the exothermicity of the reaction between years and thousands of years. However, these time values are not significant on the time scale of chemical evolution, once assumed that the surrounding conditions are maintained. This was the case of the hydrothermal vents in the bottom of Hadean/Eo-Archean oceans. In this respect, notice that the activity of the Lost City vents (Martin et al., 2008) has been estimated to be constant during the past 30000 years.

Related to the temporal question is that of the volumes of the vents (high temperature and low temperature sites) with respect to that of surrounding bulk water. Abyssal vents in Hadean/Eo-Arcehan eons should be very high in number, therefore, we could imagine a scenario of a water bulk volume containing multiple small volume hot regions.

*The question of the low concentrations necessary for SMSB.* The herein discussed SMSB occurs at very dilute concentrations. In our opinion, this is a relevant result with respect to the concentration problem in the origin of life implying aqueous media (De Duve, 1991, Russell and Martin, 2004), i.e., of the chemical evolution on Earth. The emergence of chirality would occur before the compartmentalization process to micro volumes, the primordial cell-like entities, which could be made already using enantiopure compounds.

Solubility of organic compounds in water is extremely low and in this respect one of the appealing characteristics of abyssal hydrothermal vents in origin of life scenarios is the increase of the solubility of organic compounds due to their high temperatures and/or to the presence of supercritical water. However, also in such conditions, for many significant organic compounds it is difficult to assume aqueous solutions concentrations higher than μM and it is unreasonable to expect concentrations above the mM range.

*The question of the initial chiral induction.* Although LES model discussed here is able to lead to high *ee* by the influence of an initial *ee* below the statistical fluctuations of the ideal racemic mixture (stochastic chiral sign outcome) and by so small chiral inductions as those originated by PVED (deterministic chiral sign outcome), this is probably meaningless because the SMSB discussed here can select the chiral sign of the outcome by a small initial *ee* (deterministic chiral sign of outcome) originated in other previous less sensitive processes: for example from eutectic precipitations and differential absorption phenomena between racemic and enantiopure species (Guijarro and Yus, 2009).

## IMPLICATIONS

The system described here is a closed system that differs from a more realistic scenario with respect to: 1) the actual volumes in the different temperature regions; 2) the imperfect mixing between vent and surrounding bulk water and; 3) the simplified reaction network. A better model would be that of a two-compartment system in an open system (mass and energy exchange with the surroundings). However, when bifurcation to SMSB occurs in the closed compartmentalized system, it would also occur more easily in an analogous open system. The objective of this communication is to show the possibility, in the framework of the thermodynamic feasibility, of the LES model in an abiotic scenario as an example of the possible emergence of chirality at an early stage of chemical evolution at Earth.

In our opinion the search for the conditions of coupled reaction networks implying enantioselective autocatalysis for achieving the bifurcation of SMSB is warranted. Such basic networks of SMSB probably remain in living entities nowadays, which are not only homochiral but also resistent to racemization before death. Therefore the advance in knowledge of this topic would contribute to the understanding of the conditions for life in primordial cells and in the design of synthetic primordial-cell type entities. The previous discussion together with the recent experimental results on SMSB in phase transitions (Cintas and Viedma, 2011; El-Hachemi et al., 2011; Noorduin et al., 2011; Viedma, 2005; Viedma et al. 2011) indicate that the experimental and *in silico* study of coupled reaction networks involving enantioselective autocatalysis should be undertaken under non-equilibrium conditions, either by establishing a stable temperature gradient or by the energy input to only some specific species of the system (Plasson, 2004). The temperature gradient is a scenario that would be acting during the early stages of prebiotic chemistry, but in the evolution towards more complex systems it would be substituted by the energy activation of some specific species of the system. In this respect, sulfide and phosphate bonding is an obviously reasonable acceptable hypothesis for the energy activation of certain chemical species.

Finally, it is worth noting that the basic elements of the SMSB of the LES model discussed here, are those of compartmentalization of chemical reactions and the mediating role of enantioselective enzyme-like species, both of them characteristics that are present in present day life.

## ACKNOWLEDGMENTS

The study was financed by the Spanish Government (AYA2009-13920-C01 and –C02) and forms part of the COST Action CM0703.